\documentstyle[aps,preprint,eqsecnum,epsfig]{revtex}
\input{epsf.tex}
\input{psfig.sty}
\begin{document}
\draft
\title{Logarithmic Corrections to the Equation of State in 
the SU(2)$\otimes$SU(2) Nambu -- Jona-Lasinio Model}
\author{Simon Hands}
\address{Department of Physics, University of Wales Swansea,\\
Singleton Park, Swansea, SA2 8PP, U.K.}
\author{John B. Kogut}
\address{Department of Physics,
University of Illinois at Urbana-Champaign\\
1110 West Green Street,
Urbana, IL 61801-3080, U.S.A.}
\maketitle

\begin{abstract}
We present results from a Monte Carlo simulation of the Nambu --
Jona-Lasinio model, with continuous $\mbox{SU(2)}\otimes\mbox{SU(2)}$ chiral
symmetry, in four Euclidean dimensions. Different model equations of
state, corresponding to different theoretical scenarios,  are tested
against the order parameter data. The results are sensitive to
necessary assumptions about the shape and extent of the scaling region.
Our best fits favour a trivial scenario in which the logarithmic
corrections are qualitatively similar to those predicted by the large
$N_f$ approximation. This is supported by a separate analysis of finite
volume corrections for data taken directly in the chiral limit. 
\end{abstract}
\vfill
\pagebreak

\section{Introduction}

The Nambu -- Jona-Lasinio (NJL) model \cite{Nambu} is a quantum field
theory of fermions in which the only interaction is an attractive
contact interaction between fermion and anti-fermion. Its Lagrangian
density is
\begin{equation}
{\cal L}=\bar\psi_{\alpha p}(\partial{\!\!\! /}\,+m)\psi_{\alpha p}
-{g^2\over{2N_f}}\left[(\bar\psi_{\alpha p}\psi_{\alpha p})^2-
(\bar\psi_{\alpha p}\gamma_5\vec\tau_{pq}\psi_{\alpha q})^2\right],
\label{eq:NJL}
\end{equation}
where $\vec\tau_{pq}$ are Pauli spin matrices with indices $p,q$ 
running over an SU(2) internal symmetry, and $\alpha$ runs over $N_f$
distinct fermion flavors. For bare mass $m=0$ there is an
$\mbox{SU(2)}\otimes\mbox{SU(2)}$ chiral symmetry
\begin{equation}
\psi_L\mapsto U\psi_L\;\;\;;\;\;\;\psi_R\mapsto V\psi_R,
\end{equation}
where $U$ and $V$ are independent global SU(2) rotations.
The original motivation for studying the model (\ref{eq:NJL}) is that
for coupling $g^2$ larger than some critical $g_c^2$ it gives a
qualitatively good description of dynamical chiral symmetry breaking in
strong interaction physics, with order parameter
$\Sigma\propto\langle\bar\psi\psi\rangle$. 
At the large coupling strengths required for
chiral symmetry breaking, the fermions interact by exchange of composite
scalars and pseudoscalars. In the chiral limit the pseudoscalars are
Goldstone modes, and are associated with physical pions.
More recently, NJL models have been proposed as a description of the
Higgs sector of the standard model \cite{topmode}, in which the Higgs
scalar appears as a fermion -- anti-fermion bound state, and the
Goldstone modes are absorbed into longitudinal gauge boson degrees of
freedom.

In weak coupling perturbation theory (WCPT), the NJL model is
non-renormalisable for $d>2$, being plagued by quadratic divergences
in four dimensions. However, a perturbative expansion made about the
critical point $g^2=g_c^2$ in powers of $1/N_f$ is known to be
renormalisable for $2<d<4$ \cite{RWP}\cite{HKK2}, and in $d=4$ 
suffers from only logarithmic divergences, implying that a moderate
separation between UV and IR scales can be made. Because the UV scale 
can never be completely eliminated from the theory by renormalisation
without rendering the interaction strength vanishing, the model is
{\em trivial\/} for $d=4$; it shares this property with the model more
conventionally used to describe the Higgs sector, namely the O(4)
linear sigma model, which contains elementary scalar fields. Both models
are believed trivial, and hence provide effective field theories only
for scales $\ll\Lambda$, the UV scale. At scales $\sim\Lambda$,
logarithmic scaling violations will manifest themselves, which in turn
necessitate the introduction of higher dimensional operators
corresponding to new physical information. Taking these new operators
into account, then in the sense that the number of undetermined parameters is
the same for both NJL and sigma models, the two models have equal
predictive power for the standard model \cite{Has}.

If we formulate definite lattice (ie. bare) actions based on these
models, however, an approach perhaps more familiar in
condensed matter physics, then triviality may manifest itself in distinct ways.
For both models the upper critical dimension is four, ie. the symmetry
breaking transition is described by mean-field (Landau-Ginzburg)
critical exponents with logarithmic corrections. In the sigma model, 
deviations from mean field behaviour are well-described by WCPT, since
the renormalised coupling is bounded by the fixed point value
$g_R=O(4-d)$ and is small as $d\to4$. Using the renormalisation group
\cite{Brezin}\cite{Zinn},
the following prediction for the equation of state, relating the order
parameter $\Sigma$ to the reduced coupling $t={1\over g^2}-{1\over
g_c^2}$ and the symmetry-breaking field $m$ (we retain the notation of
the NJL model) is derived:
\begin{equation}
m=At{\Sigma\over{\ln^{1\over2}\left({1\over\Sigma}\right)}}+
B{\Sigma^3\over{\ln\left({1\over\Sigma}\right)}}.
\label{eq:eossigma}
\end{equation}
Note that $\Sigma$ is measured in units of the cutoff, so that the
argument of the logarithms diverges in the continuum limit. By
contrast, symmetry breaking in the NJL model occurs for large coupling
strength, and WCPT is inapplicable. An alternative expansion parameter,
$1/N_f$, can be used \cite{EgShz}; at leading order this yields a
qualitatively different form for the equation of state \cite{KK}
\begin{equation}
m=At\Sigma+B\Sigma^3\ln\left({1\over\Sigma}\right).
\label{eq:eosnjl}
\end{equation}
The different behaviour can be traced to the fact that in the  sigma
model the scalar excitations are elementary, and the associated 
wavefunction renormalisation constant $Z$ perturbatively close to one,
whereas in the NJL model the scalars are composite, so that $Z$
vanishes in the continuum limit for $d<4$ where the model is
renormalisable, and $Z\propto1/\ln({1\over\Sigma})$ for $d=4$
\cite{EgShz}\cite{KK}; hence we expect (\ref{eq:eosnjl}) to hold
beyond leading order in $1/N_f$.

The form (\ref{eq:eossigma}) in the sigma and related scalar models
has been subject to extensive analysis via Monte Carlo simulation
(two recent examples are refs.\cite{Kenna}\cite{Gockagain}), which 
appear to provide support, although logarithmic corrections are
notoriously hard to isolate numerically. For the fermionic case, a
simulation of the $d=4$ Gross-Neveu model, with discrete chiral
symmetry, has shown support for the form (\ref{eq:eosnjl})
\cite{KKK}. The issue of which is the appropriate 
equation of state is important
for several reasons; for instance, it informs the study of the
inherently non-perturbative chiral symmetry breaking transition in
lattice $\mbox{QED}_4$. Some studies \cite{Gock1}\cite{Gock2}
have assumed a scenario of
triviality for this model based on a form similar to (\ref{eq:eossigma}),
which may not be appropriate. Another possibility is that different
lattice
implementations of the chirally symmetric four-fermi interaction,
namely the ``dual site'' formulation \cite{CER} used in \cite{KKK}
and in this study, and the ``link'' formulation used in a Monte Carlo
study of the U(1) NJL model in \cite{AliK},  may
lie in different universality classes with distinct patterns of
logarithmic scaling
violations. Studies of the corresponding lattice models in three
spacetime dimensions (``dual site'' in \cite{HKK2} and ``link''
in \cite{DelDeb}), in which the two implementations yield
radically different behaviour, support this scenario.

In this paper we extend the work of \cite{KKK} with a numerical
simulation of the $\mbox{SU(2)}\otimes\mbox{SU(2)}$ NJL model, for a
small number of flavors $N_f=2$. The main motivation for this choice of
four-fermi model, apart from the phenomenological reasons cited above,
is that it has a maximal number of Goldstone excitations, all of which
contribute to the $O(1/N_f)$ quantum corrections to the equation of
state. We hope to discriminate between the two forms (\ref{eq:eossigma})
and (\ref{eq:eosnjl}), and in particular to see if (\ref{eq:eosnjl})
persists beyond leading order in $1/N_f$. In Sec. II we outline the
lattice formulation of (\ref{eq:NJL}) using staggered fermions, and
discuss its interpretation in terms of continuum four-component spinors.
It turns out that in our simulation each lattice species describes eight
continuum flavors, and hence the $N_f=2$ model requires $N={1\over4}$
lattice flavors. This necessitates the use of a hybrid molecular
dynamics algorithm \cite{DuaneKog}\cite{GLTRS}. Extensive studies on
small systems reveal the systematic error in the algorithm to be 
$O(N^2\Delta\tau^2)$, where $\Delta\tau$ is the timestep used in the
evolution of the hybrid equations of motion; hence the systematic error
can be kept under control. Of course, we are still left with an
unquantified error due to the use of a fractional $N$, which implies a
non-local action, but have no reason to suspect there is a problem.

In Sec. III we confront data taken on a $16^4$ lattice with a range
of bare mass values in the critical region
$1/g^2\simeq0.5$, with five prototype equations of state reflecting
various theoretical prejudices, including the simplest mean field
ansatz, and a non-trivial continuum limit described by non-mean field
critical indices, 
as well as (\ref{eq:eossigma}) and (\ref{eq:eosnjl}). No form
satisfactorily fits all the data; this is because there is a finite
scaling window where the correlation length is sufficiently large for a
continuum description to apply. We show that the most successful equation
of state ansatz, in the sense of small $\chi^2$, depends on which
assumptions are made about the extent of the scaling window: if we
include all mass values from a narrow region about the critical coupling
then forms similar to (\ref{eq:eossigma}) are the most successful,
whereas if only small mass values from a wide range of couplings are
used, then (\ref{eq:eosnjl}) is the most appropriate form.

In Sec. IV we examine data taken at vanishing bare mass, which is
possible in four-fermi models formulated via the dual-site approach. Due
to the masslessness of the pion in this limit the data is subject to a
significant finite volume effect, which we attempt to account for using
a formula first derived for the sigma model \cite{Gockagain}. We find that
the variation of the finite volume correction with $g^2$ can best be
explained by assuming an equation of state (\ref{eq:eosnjl}), together
with a wavefunction renormalisation constant
$Z\propto1/\ln\left({1\over\Sigma}\right)$: in other words, it is the
finite volume corrections that seem most sensitive to the composite
nature of the scalar excitations. The distinct triviality scenarios in
NJL and sigma models \cite{KK}\cite{KKK} are rephrased in terms of the
pion decay constant $f_\pi$. In Sec. V we present brief conclusions
and directions for further work.

\section{Lattice Formulation of the Model}

The lattice action we have chosen to study is written
\begin{equation}
S_{bos}=\sum_{\alpha=1}^N\sum_{xy}
\Psi_\alpha^\dagger(x)(M^\dagger M)^{-1}_{xy}\Psi_\alpha(y)+
{{2N}\over g^2}\sum_{\tilde x}\left(\sigma^2({\tilde x})+
\vec\pi({\tilde x}).\vec\pi({\tilde x})\right),
\label{S_bos}\end{equation}
where $\Psi_\alpha$ are bosonic pseudofermion fields defined in the
fundamental representation of SU(2) on lattice sites $x$, the scalar
field $\sigma$ and pseudoscalar triplet $\vec\pi$ are defined on the
dual lattice sites $\tilde x$, and the index $\alpha$ runs over $N$
species of pseudofermion; the relation between $N$ and the number of
continuum flavors $N_f$ will be made clear presently. The fermion
kinetic matrix $M$ is the usual one for Gross-Neveu models with
staggered lattice fermions \cite{CER}\cite{HKK2}, amended to
incorporate an $\mbox{SU(2)}\otimes\mbox{SU(2)}$ chiral symmetry:
\begin{equation}
M_{xy}=\left({1\over2}\sum_\mu\eta_\mu(x)[\delta_{y,x+\hat\mu}
-\delta_{y,x-\hat\mu}]+m\delta_{xy}\right)\delta_{\alpha\beta}
\delta_{pq}
+{1\over16}\delta_{xy}\delta_{\alpha\beta}\sum_{\langle\tilde x,x\rangle}
\Biggl(\sigma(\tilde x)+i\varepsilon(x)\vec\pi(\tilde
x).\vec\tau_{pq}\Biggr),
\end{equation}
where $m$ is the bare fermion mass, the SU(2) indices $p,q$ are shown
explicitly, $\varepsilon(x)\equiv(-1)^{x_1+x_2+x_3+x_4}$, and the
symbol $\langle\tilde x,x\rangle$ denotes the set of 16 dual sites
$\tilde x$ adjacent to the direct lattice site $x$. The 3 Pauli matrices
$\tau^i$ are normalised such that ${\rm
tr}(\tau^i\tau^j)=2\delta^{ij}$.

Integration over the pseudofermions yields the following path integral:
\begin{equation}
Z\propto\int D\sigma D\vec\pi\;{\rm det}^N(M^\dagger M)\exp\left(
-{{2N}\over g^2}\sum_x(\sigma^2+\vec\pi.\vec\pi)\right).
\label{hybrid}\end{equation}
In terms of Grassmann fermion fields $\chi,\bar\chi,\zeta,\bar\zeta$,
$Z$ can be derived from the equivalent action
\begin{equation}
S=\sum_\alpha\left(\bar\chi_\alpha M[\sigma,\vec\pi]
\chi_\alpha+\bar\zeta_\alpha
M^\dagger[\sigma,\vec\pi]\zeta_\alpha\right)+{{2N}\over
g^2}\sum_x(\sigma^2+\vec\pi.\vec\pi).
\label{S}\end{equation}
The auxiliary fields $\sigma$ and $\vec\pi$ can then be integrated out
to yield an action written completely in terms of fermion fields:
\begin{eqnarray}
S_{fer}&=&\sum_\alpha\sum_{xy}\bar\chi_\alpha(x)(\partial{\!\!\!
/}\,+m)_{xy}\chi_\alpha(y)\nonumber\\
& &-{g^2\over{8N}}\sum_{\tilde x}\left[
\left({1\over16}\sum_\alpha\sum_{\langle x,\tilde x\rangle}
\bar\chi_\alpha(x)\chi_\alpha(x)\right)^2-\sum_{i=1}^3
\left({1\over16}\sum_\alpha\sum_{\langle x,\tilde x\rangle}
\bar\chi_\alpha(x)\varepsilon(x)\vec\tau\chi_\alpha(x)\right)^2\right]
\label{Sfer}\\
& &+(\chi\mapsto\zeta;\;\bar\chi\mapsto\bar\zeta;\;\partial{\!\!\! /}\,
\mapsto-\partial{\!\!\! /}\,),\nonumber
\end{eqnarray}
with $\partial{\!\!\! /}\,_{xy}$ a shorthand for the free kinetic term
for staggered lattice fermions.
So, we see that the action (\ref{S_bos}) describes $2N$ flavors of
lattice fermion, but with a flavor symmetry group ${\rm U}(N)\otimes
{\rm U}(N)$. The interaction term in (\ref{S},\ref{Sfer}) has been
normalised so that the gap equation coming from the leading order
$1/N$ approximation agrees with that derived for lattice Gross-Neveu
models having ${\rm Z}_2$ \cite{HKK2} or U(1) \cite{HKimK} chiral
symmetries.

Next we identify the global SU(2)$\otimes$SU(2)
chiral symmetry of the lattice model.
This is most manifest in the form (\ref{S}). Let ${\cal P}_e(x),{\cal
P}_o(x)$ be even and odd site projectors defined by
\begin{equation}
{\cal P}_{e/o}(x)={1\over2}(1\pm\varepsilon(x)).
\end{equation}
Then, noting that ${\cal P}_e\partial{\!\!\! /}\,{\cal
P}_o={\cal P}_e\partial{\!\!\! /}\,$ etc, 
we find that (\ref{S}) is invariant, in
the chiral limit $m\to0$, under the combined transformation:
\begin{eqnarray}
\chi\mapsto({\cal P}_eU+{\cal P}_oV)\chi\;&;&\;\bar\chi\mapsto\bar\chi
({\cal P}_eV^\dagger+{\cal P}_oU^\dagger)\nonumber\\
\zeta\mapsto({\cal P}_eV+{\cal P}_oU)\zeta\;&;&\;\bar\zeta\mapsto\bar\zeta
({\cal P}_eU^\dagger+{\cal P}_oV^\dagger)\nonumber\\
\Phi\equiv(\sigma1\kern-4.5pt1+i\vec\pi.\vec\tau)&\mapsto&V\Phi U^\dagger,
\label{SU(2)}\end{eqnarray}
where $U,V$ are independent SU(2) rotations. Note that (\ref{SU(2)}) is
a symmetry because $\Phi$ is proportional to a SU(2) matrix, and
the auxiliary potential proportional to ${\rm tr}\Phi^\dagger\Phi$.
This property does not generalise to larger unitary groups. The
symmetry (\ref{SU(2)}) is broken explicitly by a bare fermion mass,
and spontaneously by the condensates $\langle\bar\chi\chi\rangle,
\langle\bar\zeta\zeta\rangle$.

Now let us consider the continuum flavor interpretation. It is well
known that four-dimensional staggered lattice fermions can be
interpreted in terms of four flavors of Dirac fermion, which decouple
at tree level 
in the long-wavelength limit. One way of seeing this is via a
transformation to fields $q,\bar q$ defined on the sites $y$ of a lattice
of spacing $2a$ \cite{Kluberg-Stern}:
\begin{equation}
q^{\alpha a}(y)={1\over8}\sum_A\Gamma_A^{\alpha a}\chi(A;y)\;\;;
\;\;\bar q^{\beta b}={1\over8}\sum_A\bar\chi(A;y)\Gamma_A^{*\beta b},
\label{q}\end{equation}
where $\alpha,\beta=1,\ldots,4$ run over spin degrees of freedom, 
$a,b=1,\ldots,4$ over flavor, and $A$ is a four-vector with entries either
0 or 1 which identifies the corners of the elementary hypercube
associated with site $y$; each site $x$ on the original lattice is thus
mapped to a unique combination $(A;y)$. The $4\times4$ matrix 
$\Gamma_A\equiv\gamma_1^{A_1}\gamma_2^{A_2}\gamma_3^{A_3}\gamma_4^{A_4}$,
where $\gamma_\mu$ are Dirac matrices.
The action written in the form (\ref{S}) contains interaction terms of
the form $\sigma\bar\chi\chi$, $\vec\pi.\bar\chi\vec\tau\chi$ etc. In
the $q$-basis the $\sigma$ and $\pi$ fields are not all equivalent. If
we label the dual site $(x_1+{1\over2},\ldots,x_4+{1\over2})$ by 
$(A;\tilde y)$, then for $A=(0,0,0,0)$ the interaction reads
\begin{equation}
S_{int}(0;\tilde y)=\sigma(0;\tilde y)\bar q(y)(1\kern-4.5pt1\otimes
1\kern-4.5pt1)q(y)+i\vec\pi(0;\tilde y).\bar
q(y)\vec\tau(\gamma_5\otimes\gamma_5^*)q(y),
\label{int0}\end{equation}
where the first matrix in the direct product acts on spin degrees of
freedom, and the second on flavor. Equation (\ref{int0}) resembles the
continuum interaction between fermions and the auxiliary scalars,
except that the pseudoscalar current is non-singlet (but diagonalisable)
in flavor space.
However, for $A_\mu=1,A_{\nu\not=\mu}=0$, say, the interaction is more
complicated:
\begin{eqnarray}
S_{int}(&\mu&;\tilde y)=\sigma(\mu;\tilde y)\left[
\bar q(y)q(y)+{1\over2}\Delta_\mu^+\left(\bar q(y)q(y)+
\bar
q(y)(\gamma_\mu\gamma_5\otimes\gamma_5^*\gamma_\mu^*)q(y)\right)\right]
\nonumber\\
+i\vec\pi(&\mu&;\tilde y).\left[\bar
q(y)\vec\tau(\gamma_5\otimes\gamma_5^*)q(y)+{1\over2}\Delta_\mu^+
\left(\bar q(y)\vec\tau(\gamma_5\otimes\gamma_5^*)q(y)+
\bar q(y)\vec\tau(\gamma_\mu\otimes\gamma_\mu^*)q(y)\right)\right],
\end{eqnarray}
where $\Delta_\mu^+$ is the forward difference operator on the
$y$-lattice. Hence in addition to continuum-like interactions there are
extra momentum-dependent terms coupling the auxiliary fields to
bilinears which do not respect Lorentz or flavor covariance. For $A$
vectors with more non-zero entries there are still more complicated
interactions containing more derivative terms.

One might naively claim that the terms in
$\Delta_\mu\sigma,\Delta_\mu\pi$ are $O(a)$ and hence irrelevant in
the continuum limit. This ignores the fact that $\sigma$ and $\pi$ are
auxiliary fields which have no kinetic terms to suppress high momentum
modes in the functional integral. The non-covariant terms will thus
survive integration over
$\sigma,\vec\pi$ to manifest themselves in
(\ref{Sfer}). A similar phenomenon happens in lattice
formulations of the NJL model in which the interaction term is
localised on a single link \cite{AliK}\cite{Gock}. The effects of the
non-covariant terms in fact depend on the details of the model's
dynamics. If long-range correlations develop among the auxiliary
fields, ie. if an effective kinetic term is generated by radiative
corrections, then the $\Delta_\mu\Phi$ term may become irrelevant -- this
appears to be the case in two \cite{CER}\cite{Joli} and three
\cite{HKK2} dimensional models, where in each case there is a
renormalisable expansion available. In four dimensions the NJL model is
trivial: hence no continuum limit exists. Even in this case, though, we
expect correlations to develop between the auxiliary fields in the
vicinity of the continuous chiral symmetry breaking transition.
In this case the unwanted interactions would manifest themselves as
scaling violations if an interacting effective theory is
to be described \cite{Has}.

From the arguments following (\ref{Sfer}) and (\ref{q}) we now state
the relation between the parameter
$N$ and the number of continuum flavors $N_f$:
\begin{equation}
N_f=8N.
\end{equation}
The extra factor of 2 over the usual relation for four-dimensional
gauge theories arises from the impossibility of even/odd partitioning
in the dual site approach to lattice four-fermi models: 
in other words the matrix $M$ contains diagonal
terms which are not multiples of the unit matrix. In this paper we wish
to study a small number of continuum flavors $N_f=2$, which necessitates
a fractional $N=0.25$. This can be achieved using a hybrid algorithm,
which produces a sequence of configurations weighted
according to the action (\ref{hybrid})
by Hamiltonian evolution in a fictitious time
$\tau$ \cite{DuaneKog}, the fields' conjugate momenta being
stochastically refreshed at intervals. The advantage of this approach
is that in the form (\ref{hybrid}) the variable $N$ can readily be set
to a non-integer value (though there is no longer a transformation to a
local action of the form (\ref{Sfer})); the cost is that the simulation
must be run with a discrete timestep $\Delta\tau$. In principle several
values of $\Delta\tau$ must be explored, then the limit $\Delta\tau\to0$
taken. We have chosen to implement the ``R-algorithm'' of Gottlieb {\em
et al\/} \cite{GLTRS}, for which the systematic error is claimed to be 
$O(\Delta\tau^2)$. 

We have tested the R-algorithm by extensive runs on
a $6^4$ lattice with parameters $1/g^2=0.56$ and mass $m=0.01$, the
coupling being chosen to lie well into the broken symmetry phase
according to the leading order gap equation (see below). We tested
models with $N=0.25,1,3$ and 6, with timesteps
$\Delta\tau=0.4$ (for$N\leq1$), 0.2, 0.1 and 0.05. For integer $N$ we were
also able to run directly in the $\Delta\tau=0$ limit using an exact
hybrid Monte Carlo algorithm \cite{DKPR}\cite{HKK2}. We ran for 
either 20000 ($N=3,6$) or 40000 ($N=0.25,1$) Hamiltonian trajectories
between momentum refreshments, the trajectory lengths being drawn from
a Poisson distribution with mean 1.0. 

\begin{figure}[htbp]
\centerline{
\setlength\epsfxsize{300pt}
\epsfbox{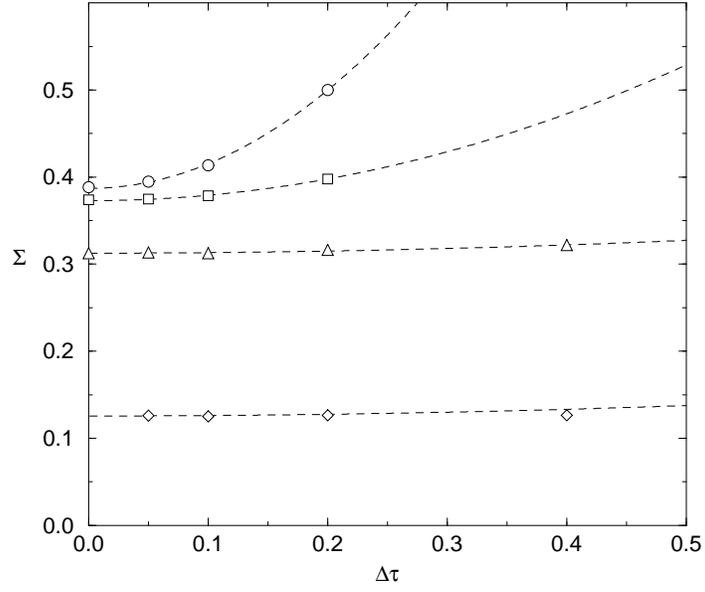}}
\caption{The condensate $\Sigma$ plotted against timestep
$\Delta\tau$ for $N=0.25$ ($\diamond$), $N=1$ ($\triangle$),
$N=3$ ($\sqcap\!\!\!\!\sqcup$) and $N=6$ ($\circ$)}
\label{fig:sigvsDt}
\end{figure}

\begin{figure}[htbp]
\centerline{
\setlength\epsfxsize{300pt}
\epsfbox{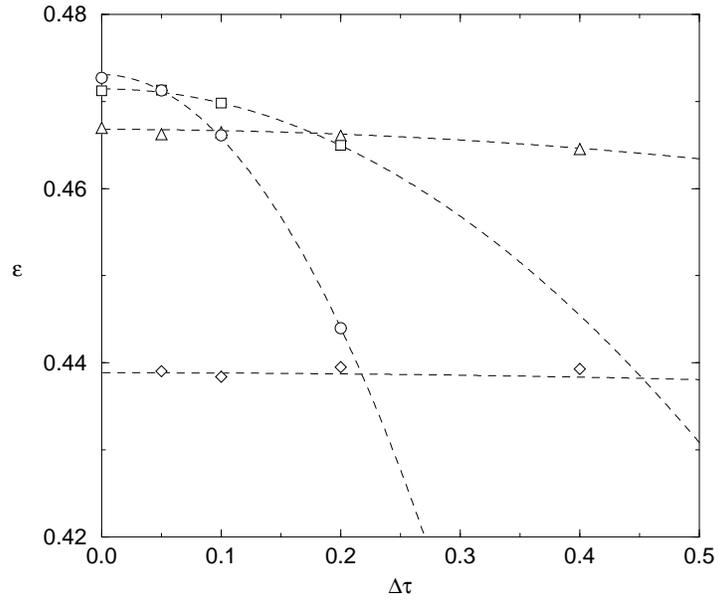}}
\caption{Same as Fig. \ref{fig:sigvsDt} but for the energy density
$\epsilon$}
\label{fig:epsvsDt}
\end{figure}

\begin{figure}[htbp]
\centerline{
\setlength\epsfxsize{300pt}
\epsfbox{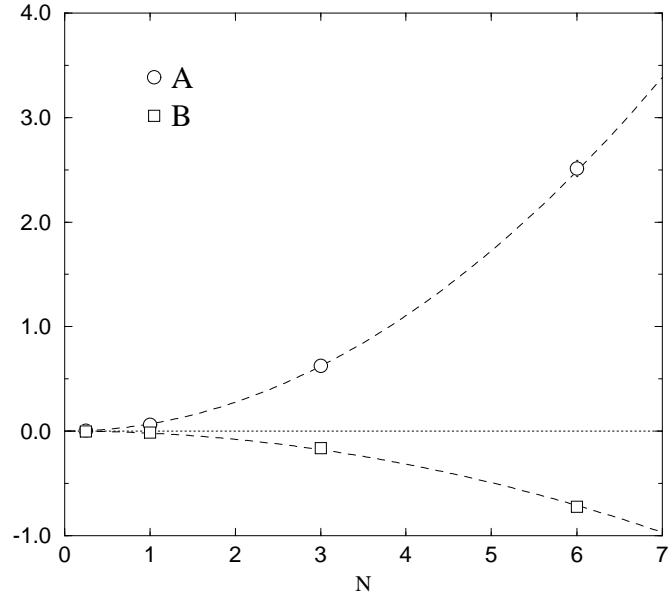}}
\caption{Behaviour of the coefficients $A(N)$ and $B(N)$}
\label{fig:ABvsN}
\end{figure}

For illustration we present results for two local quantities,
the expectation value of the scalar field $\langle\sigma\rangle\equiv\Sigma$,
and the energy density $\epsilon$ given by
\begin{equation}
\epsilon={1\over{2V}}\left\langle\sum_x
M^{-1}_{x,x+\hat4}-M^{-1}_{x,x-\hat4}\right\rangle.
\end{equation}
The results are shown in Figs. \ref{fig:sigvsDt}
and \ref{fig:epsvsDt}, together with quadratic fits of
the form
\begin{equation}
\Sigma(N;\Delta\tau)=\Sigma_0(N)+A(N)\Delta\tau^2\;\;\;;\;\;\;
\epsilon(N;\Delta\tau)=\epsilon_0(N)+B(N)\Delta\tau^2.
\end{equation}
Acceptable fits were found for all datasets except for the
$\Sigma$ data at $N=6$, where the fit was restricted to
$\Delta\tau\leq0.1$. We find good evidence that the systematic error is
indeed $O(\Delta\tau^2)$. Moreover, the coefficients $A(N)$ and $B(N)$
are themselves adequately fitted by the forms $A(N)=aN^2$, $B(N)=bN^2$
as displayed in Fig. \ref{fig:ABvsN}. This behaviour is expected from
inspection of the hybrid equations of motion \cite{DuaneKog}\cite{GLTRS}.

Since fluctuations in
the system are suppressed by powers of $1/N$, we see that for $N=0.25$
systematic errors are likely to be dwarfed by statistical ones for
reasonable simulation runs. In the work presented in the rest of this
paper we took $\Delta\tau=0.05$ just to be conservative and safe. In
fact, for very small or vanishing bare mass $m$ near the critical point
where the order parameter is very small, we also did production runs
with $\Delta\tau=0.025$ to check explicitly that the systematic errors
were smaller than the statistical errors.

The bulk of the results of this paper were generated on a $16^4$ lattice
with $N=0.25$, using masses $m$ ranging from
0.05 down to 0.0025. We have focussed our attention on the order parameter
$\Sigma$, which differs from zero in the chiral limit only
when chiral symmetry is spontaneously broken, and is simply related to
the chiral condensate:
$\Sigma={g^2\over2}\langle\bar\chi\chi\rangle$. Our
results for $\Sigma$ as a function of bare mass $m$ and
inverse coupling $1/g^2$ are presented in Tab. \ref{tab:data}. Because
the kinetic matrix $M$ contains numerically large diagonal terms, this
model is relatively cheap to simulate compared to, say, non-compact QED.
This has enabled us to accumulate a large dataset, reflected in the
relatively small statistical errors in Tab. \ref{tab:data}. 
Typical runs
which resulted in each of the entries in that table were between
15,000 and 30,000 $\tau$ units long. Such statistics are an order of
magnitude better than present day state-of-the-art lattice QCD
simulations.
The statistical error bars in the table were obtained by 
binning methods
which are particularly reliable when applied to such substantial data
sets.

In fact, because the interactions appear on the diagonal, $M$ is
sufficiently well-conditioned in the broken phase to permit simulations
directly in the chiral limit $m=0$. Of course, since spontaneous
symmetry breaking cannot occur in a finite volume $V$,
$\Sigma$ is strictly zero in these simulations; defining
\begin{equation}
\bar\Phi={1\over V}\sum_x\Phi(x),
\end{equation}
then the next
best thing to measure is
\begin{equation}
\vert\Phi\vert=\left\langle\sqrt{
\textstyle{1\over2}\mbox{tr}\bar\Phi^\dagger\bar\Phi}\right\rangle.
\label{eq:Phi}
\end{equation}
This quantity is numerically very close to $\Sigma$
extrapolated to the chiral limit, but exceeds
it in a finite volume. We will discuss the finite volume correction in
more depth in Sec. \ref{sec:fv}. Our results for 
$\vert\Phi\vert$ are given in Tab. \ref{tab:datam0}.
To monitor finite volume effects directly we did a small number of runs
on $12^4$ and $20^4$ lattices, and tabulate the results in Tabs.
\ref{tab:datam0} and \ref{tab:datafv}

\begin{figure}[htbp]
\centerline{
\setlength\epsfxsize{300pt}
\epsfbox{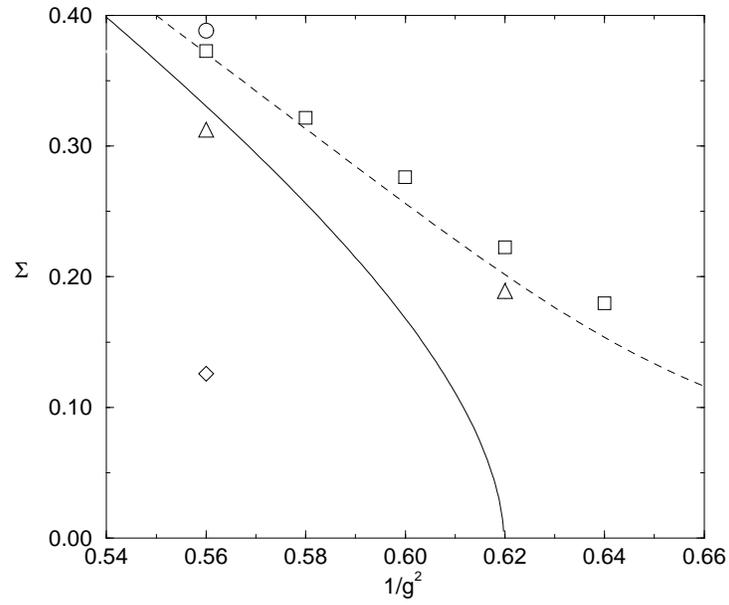}}
\caption{Gap equation predictions for $\Sigma$
for $m=0$ (solid line) and $m=0.01$ (dashed line), together with
simulation data (see caption to Fig. \ref{fig:sigvsDt})}
\label{fig:gapeqn}
\end{figure}

Finally in this section we give details of the model's gap equation. In
the large-$N$ limit, for sufficiently strong coupling $g^2$, the scalar
auxiliary $\sigma$ develops a spontaneous vacuum expectation value $\Sigma$
even in the chiral limit $m\to0$. 
To leading order in $1/N$ the relation between $\Sigma$,
the bare mass $m$ and $g^2$ is given by the lattice tadpole or gap
equation \cite{HKK2}:
\begin{eqnarray}
{1\over g^2}&=&16{{m+\Sigma}\over{\Sigma}}
\int_{-\pi/2}^{\pi/2}{{d^4k}\over(2\pi)^4}{1\over{\sin^2k_\mu+
(m+\Sigma)^2}}\nonumber\\&=&
{{m+\Sigma}\over{\Sigma}}\int_0^\infty
d\alpha\exp-\alpha\left(2+(m+\Sigma)^2\right
)I_0^4({\alpha\over2}),
\end{eqnarray}
with $I_0$ a modified Bessel function. In Fig. \ref{fig:gapeqn}
we show predictions for
$\Sigma$ as a function of $g^2$ for $m$ values
0.0, and 0.01. The $m=0$ line shows a continuous transition at 
$1/g_c^2\simeq0.62$ between a symmetric phase at weak coupling and a
broken phase at strong coupling -- the curve having the
$(g^2-g_c^2)^{1\over2}$ shape characteristic of mean field theory. For
$m>0$, $\Sigma\not=0$ for all values of $g^2$. Also shown
are simulation results from a $6^4$ lattice with $m=0.01$ for various
$N$, showing:

\begin{enumerate}

\item[(i)] that increasing fluctuations, expected to be 
$\propto1/N$, cause a suppression of $\Sigma$.

\item[(ii)] that the simulated values of $\Sigma$
exceed the gap equation predictions for $N>3$: this is in marked
contrast to the case in three dimensions, where the gap equation gives
an upper bound for all $N$ \cite{HKK2}. Perhaps the
lack of numerical accuracy of the gap equation is because the 
$1/N$ expansion is
not renormalisable in four dimensions.

\end{enumerate}

\section{Numerical Fits to the Equation of State}

In this section we report on fits to various trial forms
for the equation of state $m=f(\Sigma,1/g^2)$ 
using data for the 
order parameter
tabulated in Tab. \ref{tab:data}. 
We used the numerical package {\sc minuit} to
perform least squares fits, and in each case quote a standard error for
the fitted parameters, and the $\chi^2$ per degree of freedom, which
gives a measure of the quality of the fit.

First we list the functional forms we have examined, in increasing
order of sophistication:

\begin{enumerate}

\item[I] Mean Field (3 parameters):
\begin{equation}
m=A\left({1\over g^2}-{1\over g_c^2}\right)\Sigma+B\Sigma^3.
\label{eq:fit1}
\end{equation}
This is the simplest form, arising from a mean field treatment
neglecting fluctuations in the order parameter field.

\item[II] Power Law (5 parameters): 
\begin{equation}
m=A\left({1\over g^2}-{1\over g_c^2}\right)\Sigma^p+B\Sigma^\delta.
\label{eq:fit2}
\end{equation}
This is a more general version of I, derivable by renormalisation group
arguments from the assumption that a fixed point exists at the
transition corresponding to a diverging ratio between cutoff and
physical scales. A useful discussion is found in \cite{Zinn}. 
The index $\delta$ is the standard critical exponent,
and $p=\delta-1/\beta$, where $\beta$ is another standard exponent.

\item[III] Ladder Approximation (4 parameters):
\begin{equation}
m=A\left({1\over g^2}-{1\over g_c^2}\right)\Sigma+B\Sigma^\delta.
\label{eq:fit3}
\end{equation}
This is a restricted form of II based on the relation 
$\gamma\equiv1\Rightarrow\delta-1/\beta\equiv1$, inspired by solution
of the quenched gauged U(1) NJL model in ladder
approximation \cite{KHKD}.
\end{enumerate}

The last three fits all assume that the leading behaviour is described
by the exponents of mean field theory (ie. $\delta=3$, $p=1$), 
but with logarithmic scaling
corrections. Since these corrections introduce a dependence on some
ultraviolet scale, they imply that the renormalised theory near the
fixed point can never be made independent of the cutoff, and hence that
the continuum limit is described by a free field theory -- this is the
phenomenon of triviality. The standard scenario is derived using
renormalised perturbation theory in the context of $\mbox{O}(n)$ scalar
field theory \cite{Brezin}\cite{Zinn}. This gives rise to:

\begin{enumerate}

\item[IV] Sigma Model Logarithmic Corrections (4 parameters):
\begin{equation}
m=A\left({1\over g^2}-{1\over g_c^2}\right)
{\Sigma\over{\ln^{1\over2}\left({C\over\Sigma}\right)}}
+B{\Sigma^3\over{\ln\left({C\over\Sigma}\right)}}.
\label{eq:fit4}
\end{equation}
Here the powers of the logarithmic terms are derived on the assumption
that the effective field theory at the fixed point is the O(4) linear
sigma model, which has the same pattern of symmetry breaking as our
SU(2)$\otimes$SU(2) model. Note that we include $C$, the UV scale of
the logarithm, as a free parameter. Since the condensate data are
measured in lattice units, we expect a plausible fit should have $C$
of $O(1)$.

\item[V] Modified Sigma Model Logarithmic Corrections (5 parameters):
\begin{equation}
m=A\left({1\over g^2}-{1\over g_c^2}\right)
{\Sigma\over{\ln^{q_1}\left({1\over\Sigma}\right)}}
+B{\Sigma^3\over{\ln^{q_2}\left({1\over\Sigma}\right)}}.
\label{eq:fit5}
\end{equation}
Here we allow the powers of the logarithms to vary. This
phenomenological form has been used extensively in fits of the equation
of state of both non-compact lattice QED \cite{Gock1}\cite{Gock2}
and the U(1) NJL model \cite{AliK}. Note that
we have set the UV scale in the logarithm to 1; fits in which both $p$
and $C$ are allowed to vary are inherently unstable, since 
$\ln(C/\Sigma)=\mbox{const.}+\ln^{1/\ln C}(1/\Sigma)$, so that as
$\Sigma\to0$ the functional forms become so similar that the covariance
matrix becomes singular.

\item[VI] Large-$N$ Logarithmic Corrections (4 parameters):
\begin{equation}
m=A\left({1\over g^2}-{1\over g_c^2}\right)\Sigma
+B\Sigma^3\ln\left({C\over\Sigma}\right).
\label{eq:fit6}
\end{equation}
This form is predicted for the NJL model in the large-$N$ limit 
\cite{KK}, and was used to fit data from the four dimensional
$\mbox{Z}_2$ NJL model in \cite{KKK}. There are important qualitative
differences between form VI and the previous forms IV, V, due to the
different role of the logarithmic term -- i.e. the ``effective'' value
of the exponent $\delta$ measured at the critical coupling is larger
than the mean field value 3 for IV and V, but smaller than 3 for fit VI.
\end{enumerate}

We also tried to fit slightly modified forms of IV -- VI, eg, allowing
separate scales in the logarithms, but the results were either
unstable, or not sufficiently distinct from the six forms presented
here to be worth reporting.

As we shall see, the main issue to arise when assessing the various 
forms (\ref{eq:fit1}-\ref{eq:fit6}) is how much of the data to
include in the fit. We expect any fit only to be successful in some
finite scaling region around the transition. To see this, consider
fits using all 84 data points, shown in Tab. \ref{tab:fit1}. No fit
gives acceptable results as judged by the large $\chi^2$, although the
two 5 parameter forms II and V are both noticeably better. To proceed,
we must make some assumptions about the size and shape of the scaling
window in the $(1/g^2,m)$ plane, and truncate the dataset accordingly.
In Tab. \ref{tab:fit2} we exclude extremal inverse coupling values and
fit to $1/g^2\in[0.52,0.55]$, which still includes over half the data
in the set. The resulting fits fall into two camps. Forms I, III and
VI all yield similar large $\chi^2$ and $1/g_c^2$, and are all in effect
forced very close to the mean field form $\delta=3$, $p=1$; in the case
of VI by having the UV scale so large that the logarithmic term is
effectively constant. These fits all fail because the data in this
window does not favour $p=1$. Forms II, V and to a lesser extent IV all
allow the effective $p$ to increase from 1, e.g. for form V
$p_{eff}=1+q_1/\ln({1\over\Sigma})>1$. This enables better fits, 
which are also characterised by a smaller value for $1/g_c^2$; to see
why consider a Fisher plot of the data, in which $\Sigma^2$ is plotted
against $m/\Sigma$. The Fisher plot is designed so that the mean field
equation of state (\ref{eq:fit1}) yields curves of constant $1/g^2$
which are straight lines of uniform slope, intercepting the vertical
axis as $m\to0$ in the broken phase, the horizontal axis as $m\to0$ in
the symmetric phase, and passing through the origin as $m\to0$ for
$g^2=g_c^2$. Any departures from mean field behaviour show up as
curvature in and variation in spacing between 
lines of uniformly spaced values of $1/g^2$. 

\begin{figure}[htbp]
\centerline{
\setlength\epsfxsize{300pt}
\epsfbox{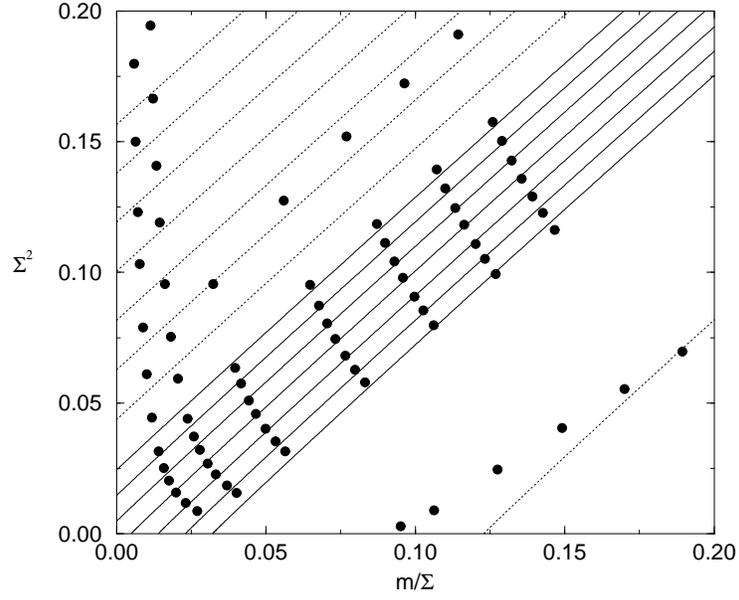}}
\caption{Fisher plot for fitting form III, $0.52\leq1/g^2\leq0.55$.
The solid lines denote the fitted data.}
\label{fig:fish3narrow}
\end{figure}

\begin{figure}[htbp]
\centerline{
\setlength\epsfxsize{300pt}
\epsfbox{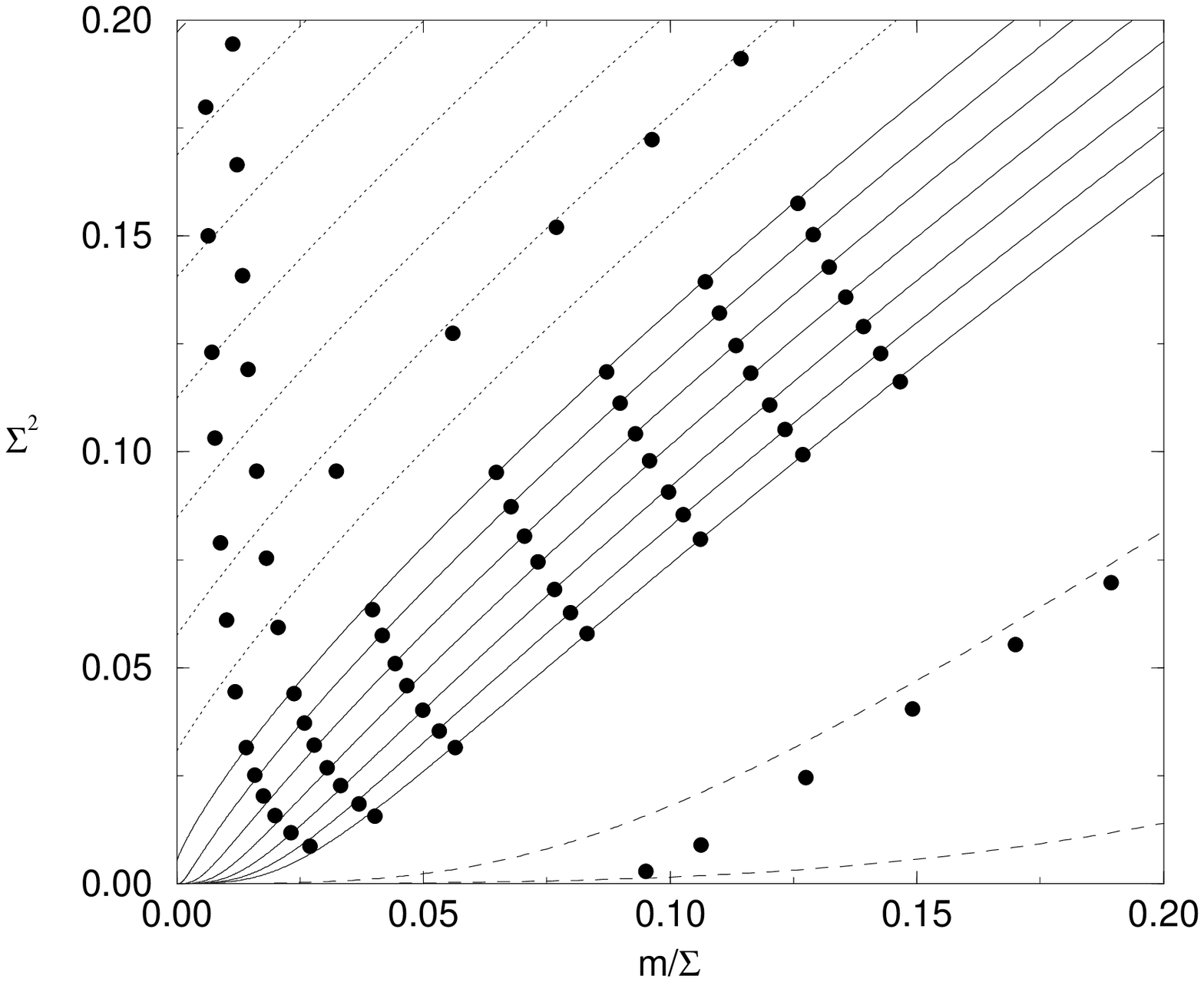}}
\caption{Fisher plot for fitting form II, $0.52\leq1/g^2\leq0.55$.}
\label{fig:fish2narrow}
\end{figure}

\begin{figure}[htbp]
\centerline{
\setlength\epsfxsize{300pt}
\epsfbox{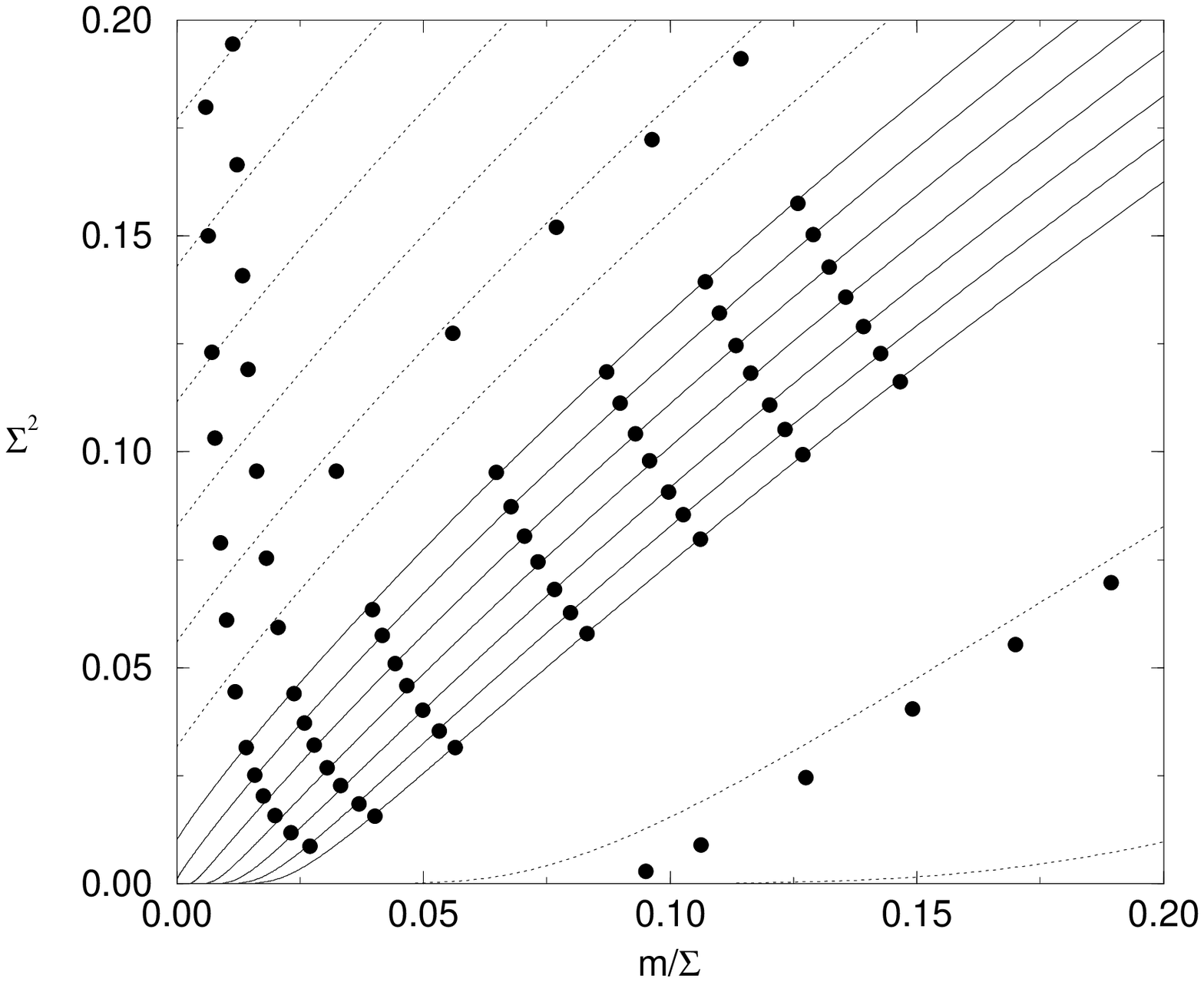}}
\caption{Fisher plot for fitting form V, $0.52\leq1/g^2\leq0.55$.}
\label{fig:fish5narrow}
\end{figure}

In Fig. \ref{fig:fish3narrow} we plot lines of constant $1/g^2$ using III, 
showing that the lines are straight and make a poor job of passing
through the data points. The plots from I and IV are almost
indistinguishable. In Fig.\ref{fig:fish2narrow} 
we show the same plot for II, and in
Fig. \ref{fig:fish5narrow}
the plot for V. These last two fits are relatively successful
because they accommodate lines of constant $1/g^2$ whose curvature
changes sign according to which phase one is in. 

Outside the fitted
window form II copes slightly better in the broken phase.
It is worth noting that forms II and V gave fits of similar quality
when applied both to simulation results
of both the U(1) NJL model \cite{AliK} and 
non-compact QED \cite{Gock2}; clearly the logarithmic
correction of (\ref{eq:fit5}) can equally well be modelled by an
effective $\delta>3$. We have checked that fits II and V were stable
under exclusion of $1/g^2=0.52,0.55$, and also under exclusion of the
low mass points $m=0.005,0.0025$.

A study of Figs. \ref{fig:fish2narrow},\ref{fig:fish5narrow}
however, indicates that fits II and V are least
satisfactory for the low mass points closest to the origin. This has
led us to explore an alternative scaling window, in which all $1/g^2$
values are kept, but high mass points are excluded. Results from just
keeping the 29 points corresponding to $m=0.0025,0.005$ are given in
Tab. \ref{tab:fit3}, and those for the 40 points obtained by also
including $m=0.01$ in Tab. \ref{tab:fit4}. The picture changes
dramatically: the $\chi^2$ values from fits III and VI are now much
smaller and comparable to those from fits II and V, though for
$m\leq0.005$ (Tab. \ref{tab:fit3}) the two 5 parameter fits still yield
the lowest values. Once $m=0.01$ points are included, fits II, III, V
and VI are virtually indistinguishable in quality.

What has happened is that the data from this window prefer an effective
value of $p$ much closer to 1, and an effective $\delta$ now less than
3. This results, e.g. in the value of $q_1$ in fit V being very small,
almost consistent with zero. The most impressive effect is that the
logarithmic correction of form VI is now acting in the correct way to
fit the data, moreover with a reasonable value for the UV scale $C$: in
fit V this manifests itself in a relatively large {\em negative\/}
value for the exponent $q_2$. The form IV, which assigns a positive value
to $q_2$, is unstable.

\begin{figure}[htbp]
\centerline{
\setlength\epsfxsize{300pt}
\epsfbox{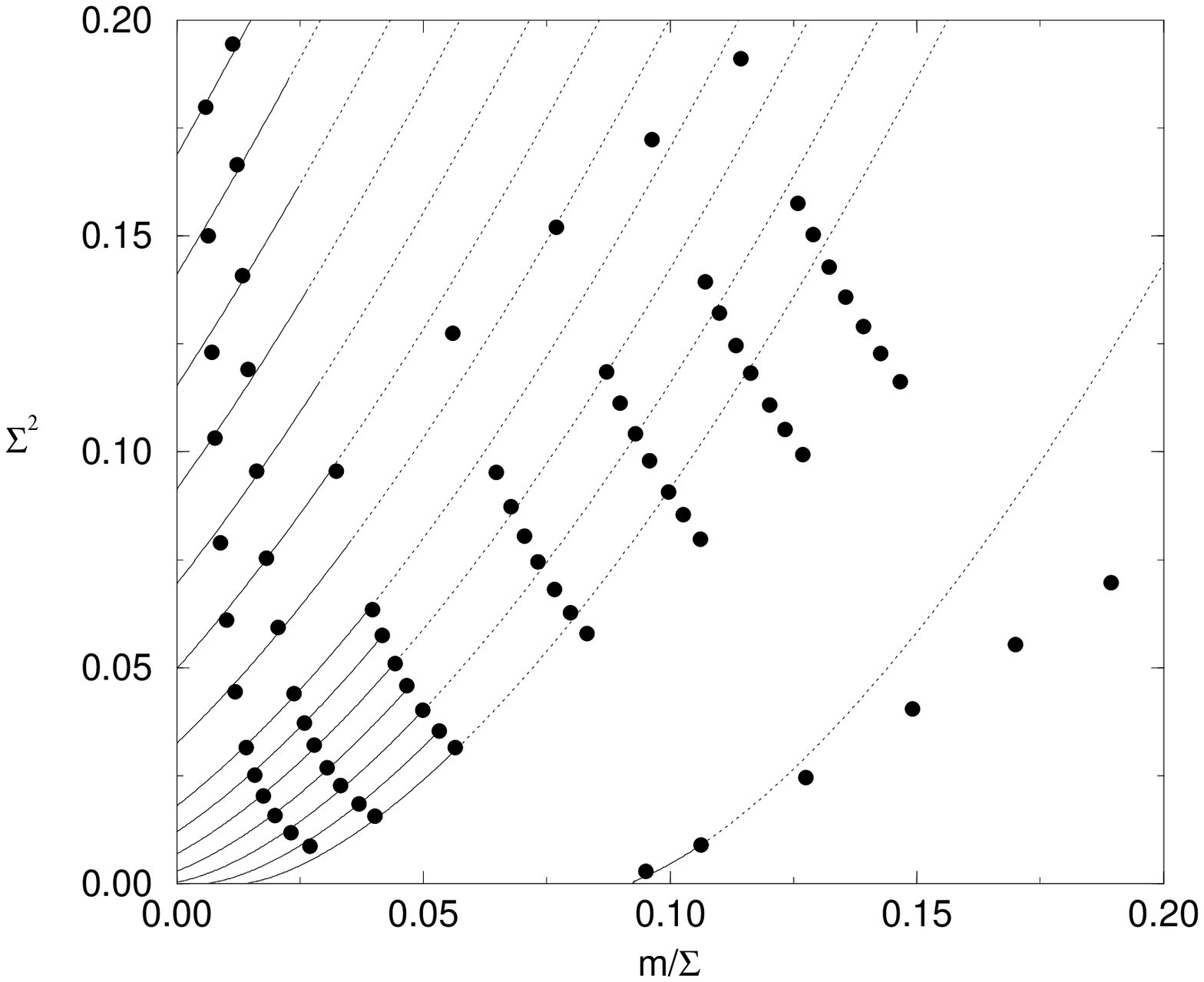}}
\caption{Fisher plot for fitting form II, $m\leq0.01$.}
\label{fig:fish2wide}
\end{figure}

\begin{figure}[htbp]
\centerline{
\setlength\epsfxsize{300pt}
\epsfbox{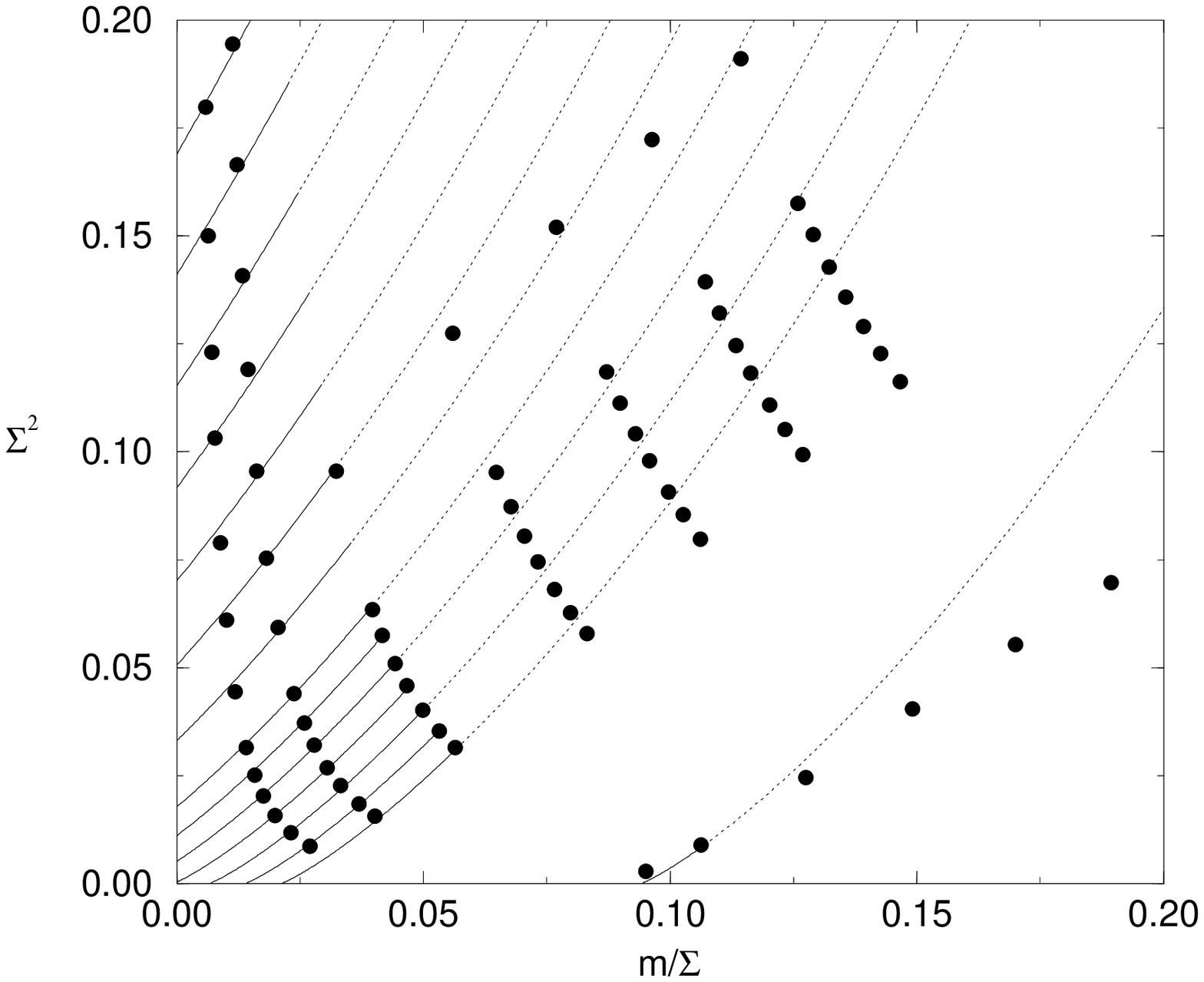}}
\caption{Fisher plot for fitting form VI, $m\leq0.01$.}
\label{fig:fish6wide}
\end{figure}

In Figs. \ref{fig:fish2wide} and \ref{fig:fish6wide}
we show Fisher plots for the forms II and VI
respectively, using the fit parameters of Tab. \ref{tab:fit4}, showing
that the fits are indeed successful over a wide range of $1/g^2$ at the
expense of missing the higher mass points. The even spacing of the
lines of constant $1/g^2$ in the broken phase is responsible for
forcing the effective value of $p$ to one.

The values of $\chi^2$ per degree of freedom for the four successful
fits are slightly too high for us to claim that they are the last word
on understanding the model's equation of state. Even though the fits
reproduce the spacing of the data in $1/g^2$, close inspection of Figs.
\ref{fig:fish2wide} and \ref{fig:fish6wide}
suggests that the variation of the data with $m$ in the broken
phase is not so well-described -- clearly more low mass data will be
needed to settle the issue, which in turn will require a quantitative
understanding of finite volume corrections. We can claim, however, 
that there appears to be a dramatic switch in the preferred form of the
equation of state according to whether the scaling window is chosen
long and thin or short and fat in the $(1/g^2,m)$ plane. In the next
section we will consider data taken directly in the chiral limit, and
find indirect evidence to support the short fat option.

\section{Finite Volume Corrections in the Chiral Limit}\label{sec:fv}

In the previous section we made no attempt to correct for finite volume
effects in the measurements of $\Sigma$. From test runs for $m=0.005$ at
$1/g^2=0.50$ on different-sized lattices, shown in Tab. \ref{tab:datafv},
finite volume effects are clearly present on a $12^4$ system, but the
difference between $16^4$ and $20^4$, though statistically significant,
is less than 2\%. We hope, therefore, that finite volume corrections
would have no impact on the qualitative conclusions of the previous
section.

Finite volume effects are numerically important, however, when
considering measurements of the quantity $\vert\Phi\vert$ made with
$m=0$, defined in (\ref{eq:Phi}) and 
shown in Tab. \ref{tab:datam0}. Naively, we expect that in a
finite system $\vert\Phi\vert$ differs from the true order parameter 
$\Sigma$ extrapolated to the chiral limit, because in the absence of a
symmetry breaking term it is impossible to disentangle fluctuations of
the order parameter field from those of the Goldstone modes; including
the effects of the latter will give $\vert\Phi\vert>\Sigma_0$, where 
$\Sigma_0$ denotes the value of $\Sigma$ in the chiral limit. 
Only in the thermodynamic limit $V\to\infty$ do we expect the Goldstone
modes to average to zero and the two
quantities to coincide. This
phenomenon also occurs in numerical studies of scalar field theory, and
has been analysed for the O(4) sigma model in \cite{Gockagain}.
Formally, $\vert\Phi\vert$ corresponds to the minimum of a ``constraint
effective potential'' in a finite volume $V$; for the O(4) sigma model
this can be calculated using renormalised perturbation theory. There is
indeed a correction due to the Goldstone modes which is intrinsically
$O(V^{-1})$. Numerically, however, a far more significant correction to
the calculation arises from the need to renormalise the model in order
to relate lattice parameters to measured quantities. To one-loop order
this results in a
relation \cite{Gockagain}
\begin{equation}
\vert\Phi\vert=\Sigma_0\left(1+B(m_RL){Z\over{\Sigma_0^2L^2}}
+O(g_R^2)\right),
\label{eq:Gock}
\end{equation}
where $m_R$, $g_R$ are the renormalised scalar mass and coupling
strength, $L$ is the linear size of the system, and $Z$ is the
wavefunction renormalisation constant, defined as the 
coefficient of
$1/p^2$ in the unrenormalised pion
propagator in the limit $p^2\to0$. 
The factor $B$ is a slowly varying function of $m_RL$ which
accounts for the difference between one-loop integrals evaluated in
finite and infinite volumes (for further background see \cite{HasLeut}).
A fit to (\ref{eq:Gock}) for magnetisation data in the sigma model is
given in Fig. 2 of \cite{Gockagain}.

For the NJL model there is no reason for renormalised perturbation
theory to apply; however since the effective theory in the broken phase
is similar to the sigma model, in the sense that it has the same light
scalar degrees of freedom, we expect a result similar to 
(\ref{eq:Gock}) to hold. We will therefore attempt to account for the
finite volume correction $\Delta=\vert\Phi\vert-\Sigma_0$ with the
formula
\begin{equation}
\Delta={{BZ}\over{\Sigma_0L^2}},
\label{eq:Delta}
\end{equation}
where the $B$ factor has been set constant, and higher order effects are
ignored. 

\begin{figure}[htbp]
\centerline{
\setlength\epsfxsize{300pt}
\epsfbox{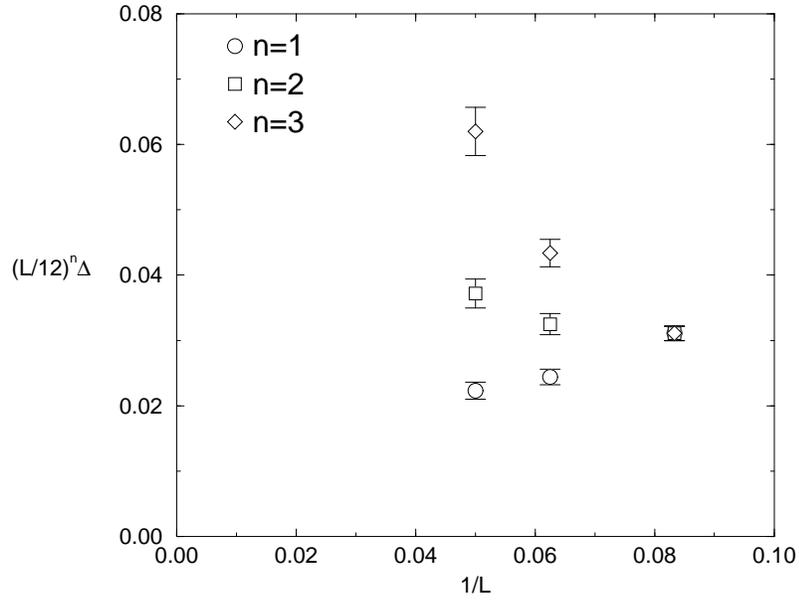}}
\caption{$(L/12)^n\Delta$ vs. $1/L$ for $1/g^2=0.48$.}
\label{fig:Delvs1onL}
\end{figure}

The first prediction of (\ref{eq:Delta}) is that 
$\Delta\propto L^{-2}$. To test this, we can compare $\vert\Phi\vert$
data taken at $1/g^2=0.48$ from three different volumes against our 
best estimate for $\Sigma_0$. From the data of Tab. \ref{tab:datam0}
we see that $\vert\Phi\vert$ decreases with $L$ as predicted.
To evaluate $\Sigma_0$, we use the two most successful fits to the
equation of state from the previous section, each one characteristic
of the two different hypotheses about the shape of the scaling window.
The narrow scaling window $0.52\leq1/g^2\leq0.55$ was best fitted by
form V (\ref{eq:fit5}) using the parameters of Tab. \ref{tab:fit2}
(the corresponding Fisher plot is shown in Fig. \ref{fig:fish5narrow});
in the chiral
limit at $1/g^2=0.48$ this form predicts $\Sigma_0=0.3341$ (we
implicitly assume here that the equation of state fits apply to the 
thermodynamic limit). This is 
already equal to the value of $\vert\Phi\vert$ on a $12^4$ lattice, and
lies above $\vert\Phi\vert$ on larger systems, making it difficult to 
see how (\ref{eq:Delta}) applies. The second estimate comes from
assuming a wide scaling window in $1/g^2$ but excluding higher mass
data. This was best described using form VI (\ref{eq:fit6}) using the 
parameters of Tab. \ref{tab:fit4} (Fig. \ref{fig:fish6wide});
the chiral limit prediction
now is $\Sigma_0=0.3030$. 
In Fig. \ref{fig:Delvs1onL} we plot $({L\over12})^n\Delta$ vs. 
$1/L$ for $n=1$, 2 and 3 and $L=12$, 16 and 20, assuming
$\Sigma_0=0.3030$. We find that $L^2\Delta$ is approximately constant,
which supports the hypothesis (\ref{eq:Delta}).

\begin{figure}[htbp]
\centerline{
\setlength\epsfxsize{300pt}
\epsfbox{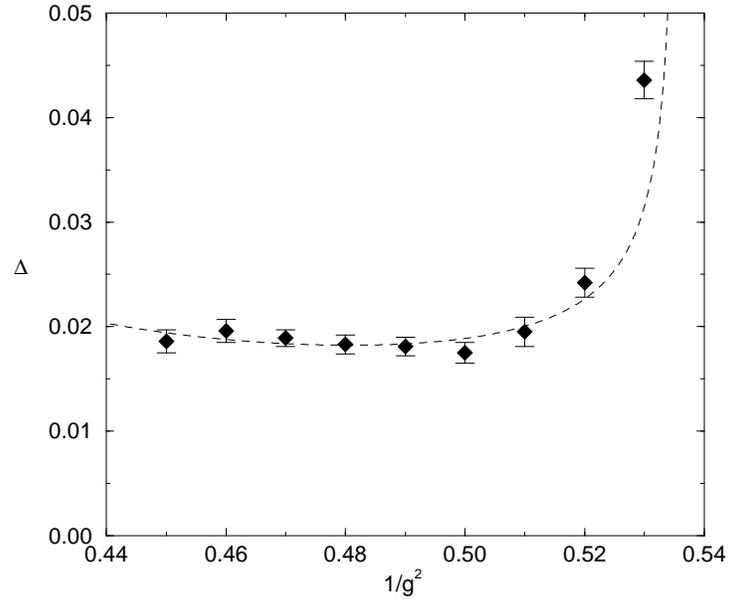}}
\caption{$\Delta$ vs. $1/g^2$ for a $16^4$ system. The line is a 
fit to (\ref{eq:fitDelta}).}
\label{fig:Delvsbeta}
\end{figure}

Next we examine the dependence of $\Delta$ on $1/g^2$, and hence
$\Sigma_0$, using data from the broken phase of the model taken from a
constant volume of $16^4$. The data of Tab \ref{tab:datam0} were used in
conjunction with the same two fits for $\Sigma_0$ to produce the two sets of
values for $\Delta$ given in Tab. \ref{tab:Delta}.
The quoted errors are the statistical errors in the measurement
of $\vert\Phi\vert$, and take no account of errors in extrapolating $\Sigma$
to the chiral limit. The values 
$\Delta_{\mbox{V}}$ obtained using form V actually change sign over the
range of $1/g^2$ explored, and clearly cannot be fitted by any relation
of the form (\ref{eq:Delta}). The values $\Delta_{\mbox{VI}}$ obtained
using form VI are plotted against $1/g^2$ in Fig. \ref{fig:Delvsbeta}. The most
interesting trend is that $\Delta_{\mbox{VI}}$ is approximately constant
for $0.45\leq1/g^2\leq0.52$, whereas $\Sigma_0$ itself falls from 
0.4115 to 0.1332 over the same range.

How do we reconcile this observation with (\ref{eq:Delta}), which 
apparently predicts $\Delta\propto1/\Sigma_0$? The answer lies in the 
field dependence of the wavefunction renormalisation constant $Z$, which
differs markedly between the ferromagnetic O(4) sigma model and the
fermionic NJL model \cite{EgShz}\cite{KK}\cite{KKK}. 
In the sigma
model, $Z$ is perturbatively close to 1 (as borne out by  the data of
\cite{Gockagain}), and hence $\Delta\propto1/\Sigma_0$. In the NJL
model, on the other hand, the large-$N_f$ approximation predicts
$Z\propto1/\ln(1/\Sigma_0)$ \cite{EgShz}, and hence
\begin{equation}
\Delta\propto{1\over{\Sigma_0\ln\left({1\over\Sigma_0}\right)}}. 
\label{eq:Delta2}
\end{equation}
We can see this using 
another argument. In both sigma and NJL models, a Ward identity plus the
assumption that both conserved current and transverse field couple
principally to the Goldstone mode (in particle physics this mode
is the pion, and the assumption is PCAC), leads to the relation
\cite{HasLeut}
\begin{equation}
Zf_\pi^2=\Sigma_0^2,
\end{equation}
where $f_\pi$ is the coupling to the axial current, or in the context of
hadronic physics the pion decay constant. We thus have 
\begin{equation}
\Delta\propto{\Sigma_0\over{(f_\pi L)^2}}.
\end{equation}
Now, in the large-$N_f$ approximation (see Appendix for an explicit
calculation),
\begin{equation}
f_\pi={{\sqrt{N_f}\Sigma_0}\over\pi}\ln^{1\over2}
\left({1\over\Sigma_0}\right),
\end{equation}
yielding once again the relation (\ref{eq:Delta2}).

Note that both scenarios predict triviality of the resulting theory.
In the NJL model, in the large-$N_f$ treatment $\Sigma_0$ is a physical
scale related to the renormalised fermion and scalar masses. The pion
decay constant in physical units is thus $f_\pi/\Sigma_0$, which
diverges as $\ln^{1\over2}(1/\Sigma_0)$ in the continuum limit
$\Sigma_0\to0$. In the sigma model, $f_\pi/\Sigma_0=Z^{1\over2}$ which
is constant in the continuum limit; however the physical scale is now
the scalar mass $m_\sigma\propto\surd{g_R}\Sigma_0$. Hence in physical
units
\begin{equation}
{{f_\pi}\over{m_\sigma}}\propto{1\over\surd{g_R}}\propto
{\ln^{1\over2}\left({1\over\Sigma_0}\right)},
\end{equation}
and again diverges in the continuum limit.

We tested the hypothesis (\ref{eq:Delta2}) with a two parameter fit of
the data of Tab. \ref{tab:Delta} to the form
\begin{equation}
\Delta={A\over{16^2\Sigma_0\ln\left({B\over\Sigma_0}\right)}}.
\label{eq:fitDelta}
\end{equation}
With all points included we find $A=1.86(13)$, $B=1.10(11)$ with
$\chi^2/\mbox{dof}$ of 4.3; however if the $1/g^2=0.53$ point is left
out the fit is much more satisfactory: $A=1.40(10)$, $B=0.82(6)$ with
$\chi^2/\mbox{dof}$ of 0.81. This fit is plotted in Fig. \ref{fig:Delvsbeta}.
In either
case it is reassuring that the UV scale of the logarithm is close to
unity. It is also interesting that the fit accounts for the slight
curvature in the plot, though this aspect is not tightly constrained by
the current error bars. The fitted curve also rises sharply to pass
close to the excluded point.

To conclude, we can account for our finite volume effects in the chiral
limit under three assumptions:

\begin{enumerate}

\item[(i)] that the equation of state fits to the $\Sigma$ data from the
$16^4$ lattice in the
previous section, when extrapolated to the chiral limit, accurately
describe the thermodynamic limit.

\item[(ii)] that the most appropriate form to extrapolate the equation of state
is form VI (\ref{eq:fit6}), using the parameters of Tab. \ref{tab:fit4}.

\item[(iii)] that the finite volume correction has the form
(\ref{eq:Delta2}).

\end{enumerate}

The latter two assumptions are consistent with logarithmic corrections
of the form advocated in \cite{KK}, which differ qualitatively from
those used to describe triviality in scalar field theory.

\section{Discussion}

It would appear that for the lattice sizes and bare masses used in this
study, which are close to state-of-the-art (though recall that
dual-site four-fermi models are relatively cheap to simulate), it is
still difficult to distinguish different patterns of triviality, or
indeed distinguish a trivial from a non-trivial fixed point, purely on
the basis of fits to model equations of state; different assumptions
about the size and shape of the scaling region result in different best
fits. Only once the analysis of Sec. IV is applied to the data taken
at zero bare mass does the form (\ref{eq:fit6}), associated with
composite scalar degrees of freedom, appear preferred. Our principal
conclusion is thus that the form (\ref{eq:eosnjl}), (\ref{eq:fit6})
is qualitatively correct beyond leading order in the $1/N_f$ expansion. 

The sensitivity of the preferred form of the equation of state to the
shape of the assumed scaling window has implications for similar
studies in related fermionic models, noteably the U(1) NJL model
studied using the ``link'' formulation \cite{AliK}, and non-compact
$\mbox{QED}_4$ \cite{Gock2}; the two recent studies cited attempt fits
based on the power law form (\ref{eq:fit2}) and the modified sigma
model form (\ref{eq:fit5}), and find the two forms difficult to
distinguish on the basis of $\chi^2$ alone (in other words, a value of
$\delta>3$ is difficult to distinguish from a logarithmic correction
with $q_2>0$). In Tab. \ref{tab:qs} the values of $q_1$ and $q_2$ are
shown for the two scaling windows used in this study, together with the
quoted fits for the two other models from \cite{AliK} and \cite{Gock2},
as well as theoretical values for the O(4) and O(2) sigma models
\cite{Brezin}\cite{Zinn}, corresponding respectively 
to the broken $\mbox{SU(2)}\otimes\mbox{SU(2)}$ symmetry in this study,
and the broken U(1) chiral symmetry of the models of \cite{AliK} and
\cite{Gock2}. Using the data of \cite{Gock2} one can examine the
stability of the fitted $q_1$ and $q_2$ with respect to exclusion of
higher mass data; the quoted value for $q_1$ appears quite stable, that
for $q_2$ less so.

An outstanding issue is whether the disparity in the fitted values of
the $q$ exponents in Tab. \ref{tab:qs}
means that the different lattice models lie in
different universality classes. The analysis presented in this paper 
suggests that current lattice simulations are unable to decide.
Unfortunately for the two models considered in \cite{AliK}\cite{Gock2}
there is no realisable way of simulating directly in the chiral limit,
and hence the analysis of Sec. IV is unavailable. A useful direction
to explore may be a simulation of the U(1) NJL model with the dual site
formulation.

\section*{Acknowledgements}
SJH is supported by a 
PPARC Advanced fellowship and would like to thank Tristram Payne 
and Luigi Del Debbio for
their help. JBK is partially supported by the National Science
Foundation under grant NSF-PHY92-00148.
\appendix
\section*{$1/N_f$ Calculation of $f_\pi$}

\begin{figure}[htbp]
\centerline{
\setlength\epsfxsize{300pt}
\epsfbox{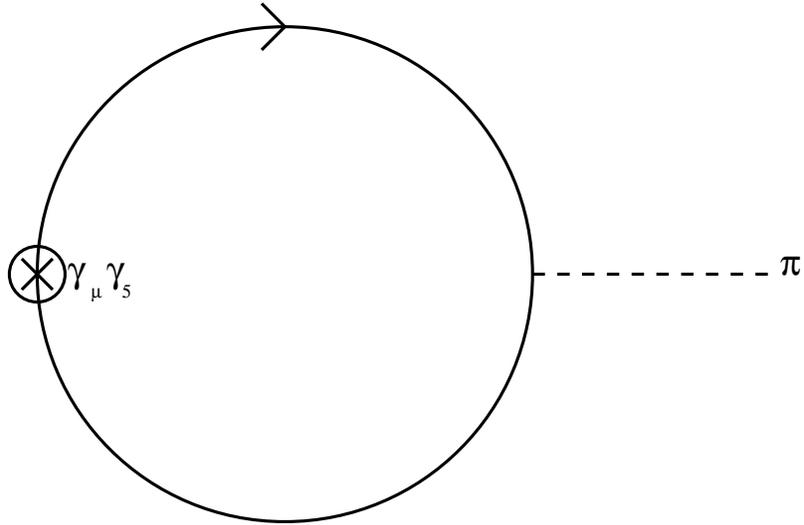}}
\caption{Leading order contribution to $f_\pi$ in the large-$N_f$
approximation.}
\label{fig:fpi}
\end{figure}

In this appendix we calculate the pion decay contant $f_\pi$,
defined by the PCAC relation
\begin{equation}
\langle0\vert A_\mu^\alpha\vert\pi^\alpha(q)\rangle\equiv
\langle 0\vert\bar\psi
i\gamma_\mu\gamma_5{{\tau^\alpha}\over2}\psi\vert\pi^\alpha(q)\rangle=iq_\mu
f_\pi,
\label{eq:PCAC}
\end{equation}
in the large-$N_f$ approximation of the NJL model.
Diagrammatically the matrix element to analyse is shown in Fig. \ref{fig:fpi}.
Note that
the external pion leg yields a normalisation factor $Z^{1\over2}$, where $Z$ is
defined by the infrared limit of the pion propagator $D_\pi$:
\begin{equation}
\displaystyle\lim_{q\to0}D_\pi(q)={Z\over q^2}.
\end{equation}
For $d<4$ $Z$ is finite. In the 
NJL model to leading order in $1/N_f$ it
is given by
\cite{HKK2}
\begin{equation}
Z={1\over g^2}{(4\pi)^{d\over2}\over{2\Gamma(2-{d\over2})}}\Sigma^{4-d}.
\label{eq:Z}
\end{equation}
The matrix element is then given by
\begin{equation}
\langle0\vert A_\mu^\alpha\vert\pi^\alpha(q)\rangle=
Z^{1\over2}N_f\times {g\over{2\sqrt{N_f}}}
\int_p\mbox{tr}\left\{{1\over{ip{\!\!\! /}\,+\Sigma}}
\gamma_\mu\gamma_5\tau^\alpha{1\over{i(p{\!\!\! /}\,+q{\!\!\!
/}\,)+\Sigma}}\gamma_5\tau^\alpha\right\},
\end{equation}
where details of Feynman rules and evaluation of $\int_p$ are also given
in \cite{HKK2}. For $d<4$ the integral can be evaluated to yield
\begin{equation}
\langle0\vert A_\mu^\alpha\vert\pi^\alpha(q)\rangle=
gZ^{1\over2}\sqrt{N_f}\times4\Sigma iq_\mu\times
{{\Gamma(2-{d\over2})}\over{(4\pi)^{d\over2}}}\int_0^1dx
{1\over{(\Sigma^2+x(1-x)q^2)^{2-{d\over2}}}}.
\end{equation}
We now take the chiral limit $\Sigma^2\gg q^2$, valid for an on-shell
pion; 
combining this step with relations (\ref{eq:PCAC}) and (\ref{eq:Z}) we find
\begin{equation}
f_\pi=2\surd2\sqrt{N_f}\Sigma^{{d\over2}-1}\sqrt{\Gamma(2-{d\over2})\over
(4\pi)^{d\over2}}.
\end{equation}
This agrees with the result found for $d=3$ in \cite{RW90}.

Finally we examine the limit $d\to4$. 
Defining $\varepsilon=4-d$, 
we have
\begin{equation}
f_\pi={{\sqrt{N_f}\Sigma}\over\pi}\sqrt{{1\over\varepsilon}-
{1\over2}\ln{\Sigma^2\over{4\pi}}-{\gamma_E\over2}+O(\varepsilon)}.
\end{equation}
In $d=4$ the model is no longer renormalisable, and we must
introduce an explicit UV cutoff, which can be set to one by
appropriate choice of units. 
On the assumption that the $1/\varepsilon$ term is then replaced
by the logarithm of the cutoff, 
we are left with the scaling prediction
\begin{equation}
f_\pi\simeq{{\sqrt{N_f}\Sigma}\over\pi}\ln^{1\over2}{1\over\Sigma}.
\end{equation}

\begin{table}[ht]
\caption{Order parameter data $\Sigma$
from a $16^4$ lattice.}
\label{tab:data}
\bigskip
\begin{tabular}{llllllll}
$1/g^2$ & $m=0.05$ & $m=0.04$ & $m=0.03$ & $m=0.02$ & $m=0.01$ &
$m=0.005$ & $m=0.0025$ \\
\hline
\hline
0.45 & 0.5503(3) & 0.5344(3) & 0.5149(3) & 0.4919(4) & 0.4611(9) 
     & 0.4411(9) & 0.4241(13) \\
0.46 &   --      &    --     &    --     &    --     &    --
     & 0.4080(11)& 0.3873(18) \\
0.47 &   --      &    --     &    --     &    --     &    --
     & 0.3753(10)& 0.3509(19) \\
0.48 &   --      &    --     &    --     &    --     &    --
     & 0.3452(11)& 0.3213(15) \\
0.49 &   --      &    --     &    --     &    --     &    --
     & 0.3091(12)& 0.2809(21) \\
0.50 & 0.4371(4) & 0.4152(5) & 0.3899(6) & 0.3570(7) & 0.3091(7)
     & 0.2746(8) & 0.2470(18) \\
0.51 &   --      &    --     &    --     &    --     &    --
     & 0.2435(8) & 0.2107(19) \\
0.52 & 0.3970(3) & 0.3733(5) & 0.3443(6) & 0.3086(5) & 0.2520(5)
     & 0.2097(9) & 0.1774(20) \\
0.525& 0.3877(3) & 0.3636(3) & 0.3337(3) & 0.2954(3) & 0.2398(4)
     & 0.1928(8) & 0.1583(17) \\
0.53 & 0.3780(3) & 0.3530(3) & 0.3228(3) & 0.2836(5) & 0.2257(7)
     & 0.1789(8) & 0.1425(19) \\
0.535& 0.3686(2) & 0.3438(4) & 0.3129(3) & 0.2730(5) & 0.2142(6)
     & 0.1637(9) & 0.1253(20) \\
0.54 & 0.3592(2) & 0.3329(2) & 0.3011(3) & 0.2611(7) & 0.2004(6)
     & 0.1506(11)& 0.1081(16) \\
0.545& 0.3505(3) & 0.3243(3) & 0.2923(3) & 0.2504(3) & 0.1879(5)
     & 0.1355(10)& 0.0926(20) \\
0.55 & 0.3409(3) & 0.3152(3) & 0.2825(3) & 0.2405(7) & 0.1773(9)
     & 0.1246(13)&    --      \\
0.60 & 0.2641(3) & 0.2353(3) & 0.2012(6) & 0.1568(7) & 0.0941(5)
     & 0.0526(10)&    --      \\
0.70 & 0.1652(3) & 0.1396(3) & 0.1112(4) & 0.0784(4) & 0.0403(3)
     & 0.0207(4) &    --      \\
\end{tabular}
\end{table}

\begin{table}[ht]
\caption{Measured values of $\vert\Phi\vert$
for $m=0$ from various lattice sizes.}
\label{tab:datam0}
\bigskip
\begin{tabular}{llll}
$1/g^2$ & $12^4$ & $16^4$ & $20^4$ \\
\hline
\hline
0.45 & -- & 0.4301(11) & -- \\
0.46 & -- & 0.3955(11) & -- \\
0.47 & -- & 0.3587(8) & -- \\
0.48 & 0.3341(11) & 0.3213(9) & 0.3164(8) \\
0.49 & -- & 0.2830(10) & -- \\
0.50 & -- & 0.2424(10) & -- \\
0.51 & -- & 0.2013(14) & -- \\
0.52 & -- & 0.1574(14) & -- \\
0.53 & -- & 0.1149(18) & -- \\
\end{tabular}
\end{table}

\begin{table}[ht]
\caption{Order parameter $\Sigma$ 
for $m=0.005$, $1/g^2=0.50$ from various lattice sizes.}
\label{tab:datafv}
\bigskip
\begin{tabular}{lll}
 $12^4$ & $16^4$ & $20^4$ \\
\hline
\hline
0.2643(14) & 0.2746(8) & 0.2786(7) \\
\end{tabular}
\end{table}
\vfill\eject

\begin{table}[ht]
\caption{Equation of state fits using all data (84 points).}
\label{tab:fit1}
\bigskip
\begin{tabular}{llllllllr}
& $A$ & $B$ & $1/g_c^2$ & $\delta$ & $p$ &
$\displaystyle\biggl\{{q_1\atop q_2}$ & $C$ & $\chi^2$/dof  \\[5pt]
\hline
\hline
 I  & 1.8608(17) & 0.8443(10) & 0.52812(6)  & --         & -- 
    & -- & -- &  559.2 \\
II  & 2.3757(74) & 0.7636(27) & 0.53941(25) & 2.6901(68) & 1.1978(21)
    & -- & -- &   73.0 \\
III & 1.7974(18) & 0.5810(10) & 0.55759(27) & 2.2237(40) &  --
    & -- & -- &   183.4 \\
IV  & 2.884(13)  & 1.497(18)  & 0.52006(9)  & --  & --
    & -- & 2.850(59) &   385.7 \\
V   & 1.9669(26) & 0.9590(22) & 0.53311(14) & --  & --
    & $\displaystyle{0.3276(30)\atop\!\!\!\!-0.0592(52)}$ & -- & 39.7 \\
VI  & 1.7877(17) & 0.6577(33) & 0.54635(12) & --  & -- 
    & --  & 2.004(12) & 182.6 \\
\end{tabular}
\end{table}

\begin{table}[ht]
\caption{Equation of state fits using data for $0.52\leq1/g^2\leq0.55$
(48 points).}
\label{tab:fit2}
\bigskip
\begin{tabular}{llllllllr}
& $A$ & $B$ & $1/g_c^2$ & $\delta$ & $p$ &
$\displaystyle\biggl\{{q_1\atop q_2}$ & $C$ & $\chi^2$/dof  \\[5pt]
\hline
\hline
 I  & 1.8152(53) & 0.9616(13) & 0.53309(7)  & --         & -- 
    & -- & -- &  45.0 \\
II  & 3.845(71) & 1.309(20) & 0.52224(44) & 3.50(2) & 1.59(1)
    & -- & -- &   3.1 \\
III & 1.8160(53) & 0.9943(89) & 0.53228(22) & 3.044(11) &  --
    & -- & -- &   45.0 \\
IV  & 3.285(58)  & 2.64(11)  & 0.5274(2)  & --  & --
    & -- & 7.13(82) &   32.0 \\
V   & 2.160(10) & 0.8499(53) & 0.52562(32) & --  & --
    & $\displaystyle{0.786(19)\atop0.372(17)}$ & -- & 2.5 \\
VI  & 1.8146(53) & 0.0258(5) & 0.53359(8) & --  & -- 
    & --  & 0.70(45)$\times10^{16}$ & 45.0 \\
\end{tabular}
\end{table}

\begin{table}[ht]
\caption{Equation of state fits using data for $m\leq0.005$
(29 points).}
\label{tab:fit3}
\bigskip
\begin{tabular}{llllllllr}
& $A$ & $B$ & $1/g_c^2$ & $\delta$ & $p$ &
$\displaystyle\biggl\{{q_1\atop q_2}$ & $C$ & $\chi^2$/dof  \\[5pt]
\hline
\hline
 I  & 1.136(13) & 0.4907(73) & 0.51910(33)  & --         & -- 
    & -- & -- &  41.0 \\
II  & 1.099(35) & 0.312(11) & 0.53781(83) & 2.226(26) & 0.921(10)
    & -- & -- &   1.5 \\
III & 1.367(19) & 0.3992(64) & 0.53779(83) & 2.351(19) &  --
    & -- & -- &   3.4 \\
IV  & &  & & no fit found  & 
    & &  &   \\
V   & 1.205(33) & 0.633(18) & 0.53157(54) & --  & --
    & $\displaystyle{\!\!\!\!-0.118(25)\atop\!\!\!\!-0.707(34)}$ & -- & 2.4 \\
VI  & 1.354(18) & 0.421(16) & 0.53334(52) & --  & -- 
    & --  & 2.01(8) & 2.9 \\
\end{tabular}
\end{table}

\begin{table}[ht]
\caption{Equation of state fits using data for $m\leq0.01$
(40 points).}
\label{tab:fit4}
\bigskip
\begin{tabular}{llllllllr}
& $A$ & $B$ & $1/g_c^2$ & $\delta$ & $p$ &
$\displaystyle\biggl\{{q_1\atop q_2}$ & $C$ & $\chi^2$/dof  \\[5pt]
\hline
\hline
 I  & 1.403(6) & 0.6349(43) & 0.52297(18)  & --         & -- 
    & -- & -- &  72.4 \\
II  & 1.382(17) & 0.3998(68) & 0.54149(61) & 2.267(18) & 0.9731(53)
    & -- & -- &   3.9 \\
III & 1.465(7) & 0.4316(35) & 0.54019(83) & 2.334(12) &  --
    & -- & -- &   4.5 \\
IV  & &  & & no fit found  & 
    & &  &   \\
V   & 1.476(13) & 0.7837(73) & 0.53391(32) & --  & --
    & $\displaystyle{0.016(11)\atop\!\!\!\!-0.553(18)}$ & -- & 4.6 \\
VI  & 1.459(7) & 0.474(10) & 0.53546(31) & --  & -- 
    & --  & 1.95(5) & 4.0 \\
\end{tabular}
\end{table}

\begin{table}[ht]
\caption{Values of the quantity $\Delta$ for the two fits to the
thermodynamic and chiral limit discussed in the text.}
\label{tab:Delta}
\bigskip
\begin{tabular}{lrrr}
 $1/g^2$ & $\Delta_{\mbox{V}}$ & $\Delta_{\mbox{VI}}$ & error \\
\hline
\hline
0.45 & -0.0326 & 0.0186 & 0.0011\\
0.46 & -0.0253 & 0.0196 & 0.0011\\
0.47 & -0.0195  & 0.0189 & 0.0008 \\
0.48 & -0.0128  & 0.0183 & 0.0009\\
0.49 & -0.0045 & 0.0181 &  0.0010\\
0.50 & 0.0058  & 0.0175 &  0.0010\\
0.51 & 0.0233  & 0.0195 & 0.0014\\
0.52 & 0.0568  & 0.0242 & 0.0014\\
0.53 & 0.1149  & 0.0436 & 0.0018\\
\end{tabular}
\end{table}

\begin{table}[ht]
\caption{Values of the $q$ exponents for the various models discussed
in Sec. V.}
\label{tab:qs}
\bigskip
\begin{tabular}{lllllll}
& O(4) & O(2) & SU(2) NJL 
& SU(2) NJL 
& U(1) NJL &
$\mbox{QED}_4$  \\
&&&(Tab. \ref{tab:fit2})&(Tab. \ref{tab:fit4}) & (Ref. \cite{AliK})&
Ref. \cite{Gock2}) \\
\hline
\hline
 $q_1$  & 0.5 & 0.4 & 0.786(19)  & 0.016(11)  & 0.36(11) & 0.485(7) \\ 
 $q_2$  & 1   & 1   & 0.372(17)  & $\!\!\!\!-0.553(18)$ & 0.84(1)  & 0.324(15)\\
\end{tabular}
\end{table}
\clearpage

\end{document}